\documentclass[graybox,nosecnum]{svmult}
\usepackage{natbib}
\usepackage{mathptmx}       
\usepackage{helvet}         
\usepackage{courier}        
\usepackage{makeidx}         
\usepackage{multicol}        
\usepackage[bottom]{footmisc}
\usepackage{hyperref}        
\usepackage{soul}            
\hypersetup{colorlinks=true,urlcolor=blue}
\makeindex             

\usepackage{graphicx}
\graphicspath{{./figures/}}
\usepackage{color}
\usepackage{amssymb}

\newcommand{\mygamma}{gamma}
\newcommand{\gammaRay}{\mygamma\ ray}
\newcommand{\gammaRays}{\mygamma\ rays}
\newcommand{\gammaRayHyph}{\mygamma -ray}
\newcommand{\fermi}{\emph{Fermi}}

\newcommand{\techname}[1]{\texttt{#1}}
\newcommand{\Psix}{\techname{Pass6}}
\newcommand{\Pseven}{\techname{Pass7}}
\newcommand{\Peight}{\techname{Pass8}}

\newcommand{\rl}{$X_0$}
\newcommand{\um}{$\mu$m}
\newcommand{\us}{$\mu$s}
\newcommand{\uW}{$\mu$W}
\newcommand{\cmsq}{cm$^2$}
\newcommand{\isreq}{$\checkmark$}
\newcommand{\isexc}{$\times$}

\newcommand{\figref}[1]{Fig.~\ref{fig:#1}}
\newcommand{\tabref}[1]{Table~\ref{tab:#1}}

\definecolor{mycolor}{rgb}{1.0,0.4,0.4}
\newcommand{\mydate}[3]{#2 #3, #1}

\begin{document}
\title*{The Fermi Large Area Telescope}
\author{Riccardo Rando}
\institute{R. Rando \at
              University of Padova and I.N.F.N. Padova\\
              via Marzolo 8, I-35131 Padova, Italy\\
              \email{riccardo.rando@pd.infn.it}
}


\maketitle

\abstract{
  The Large Area Telescope, the primary instrument on the Fermi Gamma-ray Space Telescope, is an imaging, wide field-of-view \gammaRayHyph\ telescope. After many improvements to the data acquisition and event analysis procedures, it now covers the broad energy range from $\sim 20$~MeV to $\sim  2$~TeV. After more than 13 years of operation since its launch in \mydate{2008}{June}{11}, it has provided the best-resolved and deepest portrait of the \gammaRayHyph\ sky. In this chapter we review the design of the instrument, the data acquisition system, calibration, and performance.
}

\section{Keywords} 
gamma-ray telescope, calibration, silicon microstrip detector, electromagnetic calorimeter, plastic scintillator

\section{Introduction}

The birth of \emph{multi-wavelength} astronomy dates back to the 1960s, when radio astronomy reached maturity.
Most of the electromagnetic (EM) spectrum, however, was still out of reach due to atmospheric absorption outside the radio and optical windows. To cover most of the infrared band, X rays and soft \gammaRays\ it was necessary to wait until the early space missions in the 1960s and 1970s. In the 1980s the first Imaging Atmospheric Cherenkov Telescopes (IACT) were built on the ground to reach even higher energies. Almost simultaneously instruments appeared capable of astronomical observations with probes other than photons: the observations of cosmic neutrinos signaled the birth of \emph{multi-messenger} astrophysics. For a review see e.g. \citet{multimessenger,multimessenger2}. 

The Large Area Telescope (LAT) on the Fermi Gamma-ray Space Telescope (\fermi) was designed to cover, with excellent performance, the energy range from a few MeV to several hundred GeV, overlapping with ground-based IACT observatories. Deployed at a very favorable time, it plays a major role in the multi-messenger revolution currently underway.

In this chapter we describe the design, operation and performance of the LAT instrument. The interested reader can find a description of the \fermi\ mission and an overview of the scientific results in the dedicated chapter of this Handbook. 

Firstly, we give a brief summary of the topic of \gammaRayHyph\ astrophysics from space, and how the scientific requirements and the technological advances shaped the design of the instrument. Then we describe in greater detail the three main subsystems of the LAT (tracker, calorimeter and anticoincidence detector), how the data acquisition system operates and how data are processed. We give some details on the operation of the LAT, with particular attention to the procedures required to keep the telescope in good health. Lastly, we describe the calibration of the detector and the scientific performance.

\section{\textit{A space-based MeV--GeV \gammaRayHyph\ observatory}}

Photons with energies above a few 10's of MeV interact almost exclusively through the process of electron--positron pair production. Since the original photon disappears in the interaction, its properties must be derived from measurements performed on the two daughter particles. Standard optical approaches (reflection and refraction) are not applicable, and the photons are too penetrating for collimators. An MeV--GeV telescope is, in fact, a high energy physics (HEP) detector, based on technologies developed for the use at accelerator facilities. Additional constraints are imposed by the need to place the detector in orbit, to escape the opacity of the atmosphere to high-energy photons (short observations can be performed from a balloon). 

Looking at the history of \gammaRayHyph\ observatories in space, it is apparent how much of the progress in the field is related to the technological advancement of HEP particle detectors \citep{history,egret-to-lat}.
In the pioneering years (1960s) space-based \gammaRayHyph\ detectors were typically based on a stack of scintillators, restricting the acceptance to a small angle in lieu of measuring the direction of the incoming photons \citep{oso3}. The later generation of instruments (1970s) included a gas-filled spark chamber, where the original direction of the photon could be reconstructed from the ionization tracks left by the secondaries, leading to a great increase in field of view (FOV) and sensitivity \citep{sas2,cosb}. The spark chamber was at the core of the very successful EGRET instrument, on board the Compton Gamma-Ray Observatory (CGRO) that operated from 1991 to 2000 \citep{egret}.  

In the 1990s the most important development in particle tracking since EGRET was the advent of large-area silicon strip trackers, so it was natural to design a successor around them. The silicon (micro-)strip  detector (SSD) was developed in 1980 \citep{hejne80} and used with success in major HEP facilities all around the world. The advantages with respect to the existing technologies were many: very good spatial resolution ($\geq 50$~\um), excellent signal-to-noise ratio, good radiation hardness, robustness, absence of consumables, self triggering, etc. In parallel to tracking detectors, Application Specific Integrated Circuits (ASICs) were undergoing a rapid evolution from their appearance at the end of the 1960s, enabling the readout of an unprecedented number of channels ($\sim 1$ million) with excellent performance and low power consumption. Manufacturing and design techniques were developed to enable ASICs to withstand severe radiation environments. SSDs and ASICs have been instrumental to the success of the LAT.

Towards the end of the EGRET mission, the general constraints for the next space-based \gammaRayHyph\ observatory were defined in a two-step process. Firstly, the major science targets were identified, including Active Galactic Nuclei, isotropic background radiation, Gamma Ray Bursts (GRBs), sites of cosmic ray acceleration, neutron stars and black holes, dark matter. Secondly, an estimate of the performance necessary to reach the aforementioned science goals was drafted \citep{scireq-site}. NASA released an Announcement of Opportunity in August 1999, detailing the baseline scientific objectives and soliciting proposals for the development of a large-area \gammaRayHyph\ telescope  \citep{ao-site} originally called Gamma-ray Large Area Space Telescope (GLAST). 

The proposal describing the LAT as we know it was submitted in November 1999: a design based on ``(i) a precision tracker, based on proven Silicon-strip detector technology, (ii) a finely segmented Cesium Iodide calorimeter for energy measurement, and (iii) a segmented anticoincidence that covers the tracker'' \citep{resp-ao-site}. 

Let us see in some detail how the scientific and operational constraints defined the main characteristics of the instrument.

\begin{itemize}
\item The large energy coverage, up to hundreds of GeV, sets a lower limit on the thickness, in particular on the mass of the calorimeter, in order to contain a large part of the induced electromagnetic shower. 
\item An anticoincidence shield, to veto the abundant charged particles found in space, thick enough to ensure a high detection efficiency but transparent to \gammaRays.
\item A massive calorimeter near a veto scintillator requires a way to alleviate the self-veto, caused by the low energy photons produced in the calorimeter in high-energy events going through the anticoincidence. Segmentation addresses the issue, making it possible to correlate the shower development inside the instrument with the location of the energy depositions in the surrounding veto detector.

\item For a full-sky observatory a squat aspect ratio is preferred, since the geometrical shape defines the FOV. On the other hand, a time-of-flight detector between tracker and calorimeter is very useful for trigger and background rejection, but requires a tall instrument. The choice of a self-triggering tracker made it possible to do without.
\item The size of the instrument was limited by the available launcher: for the LAT, the Delta-II ``Heavy'' launch vehicle constrained the lateral size to $\sim 2$~m and the mass to $\sim 3,000$~kg.
\item High efficiency and excellent spatial resolution in the tracker were considered key parameters. A Silicon tracker having the necessary thickness would have been too costly and with too many readout channels: conversion foils of high-Z material were interleaved, balancing the conversion efficiency and the loss of resolution due to multiple scattering in the foils. 
\item The design downlink budget of $<300$~kbps limited the number of readout channels and the number of bits available per channel, the average trigger rate and mandated the presence of a powerful background filter on board.
\item The required mission time of 5 years (with a goal of 10 years) required no consumables, substantial radiation hardness, and a highly modular structure to limit the impact of any damage or failure.
\end{itemize}

\begin{figure}[htbp]
  \centering
  \includegraphics[scale=1]{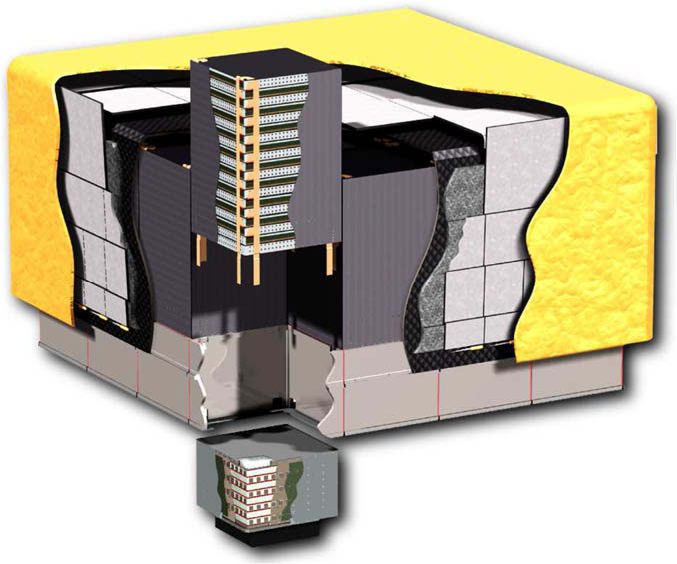}
  \caption{LAT cutout view, from \citet{tkr-perf-2-years}. Exterior, cut: thermal blanket, micro-meteoroid shield and tiled ACD (top), support grid (bottom). Inside: the 16 towers; the foremost tower is separated in the TKR module on top and CAL module on the bottom, with tower electronics barely visible below.}
  \label{fig:lat}
\end{figure}

\begin{figure}[htbp]
  \centering
  \includegraphics[width=0.45\textwidth]{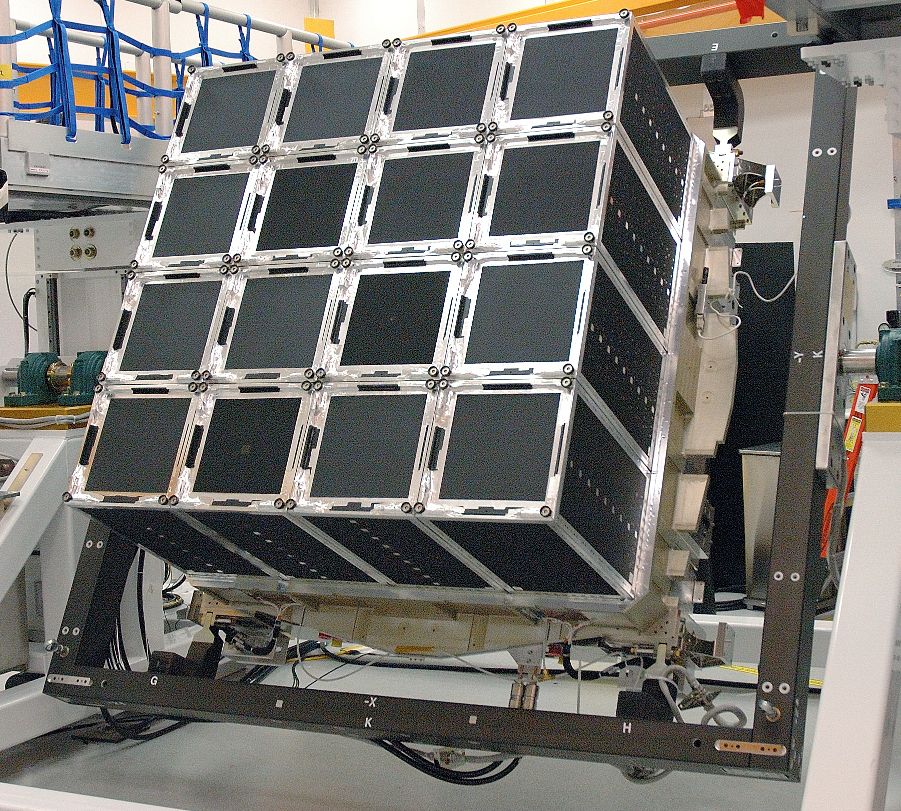}
  \includegraphics[width=0.4\textwidth]{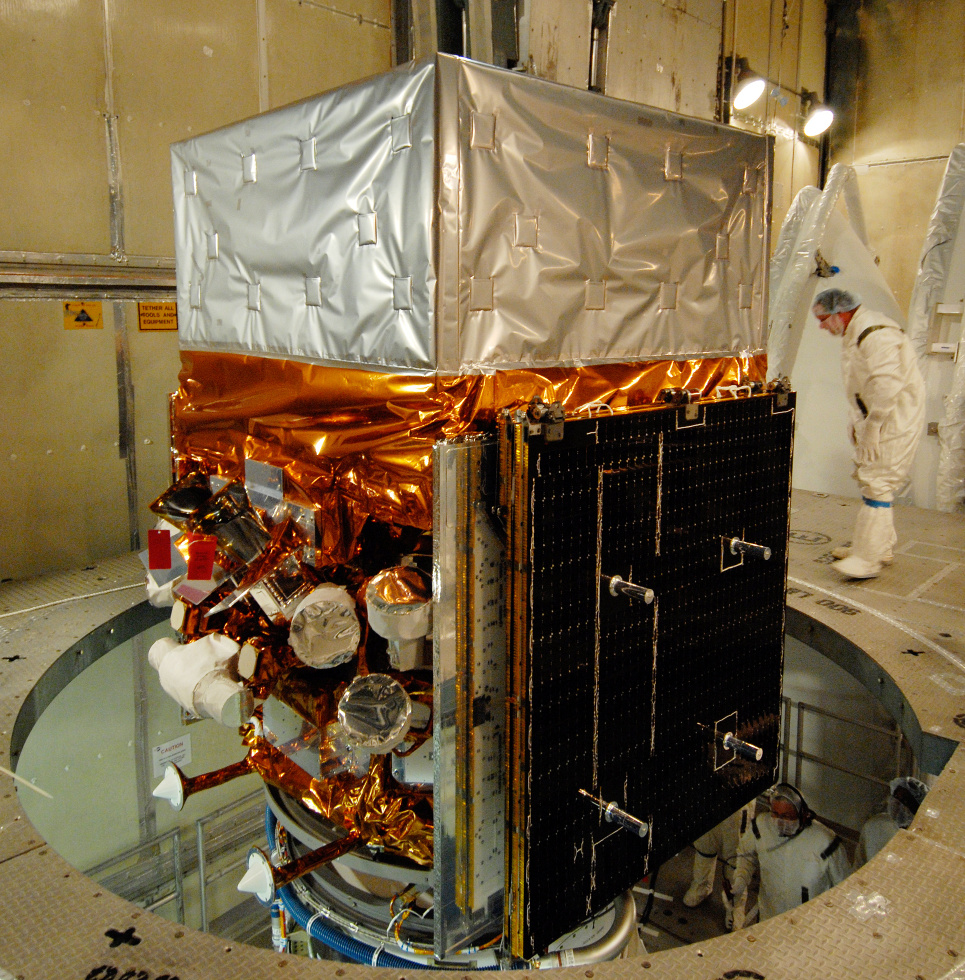}
  \caption{Left: the LAT during integration before the ACD was installed, showing the top of the 16 towers. Right: the LAT (top) on the \fermi\ spacecraft (bottom). In the left side of the spacecraft, part of the GBM is visible: 3 orthogonal NAI detectors near the corner next to 1 white-colored BGO detector. From \citet{instr10y}, \copyright\ AAS.}
  \label{fig:lat-real}
\end{figure}

The final design divided the instrument into 16 identical towers, each containing a Silicon tracker on top of a Cesium Iodide (CsI) calorimeter and including the necessary readout electronics. All towers, arranged in a $4\times 4$ array on the support structure, were enclosed in the segmented plastic anti-coincidence detector (ACD), and wrapped into a micrometeroid shield and thermal blanket. An artist's cutout view is given in \figref{lat}, to be compared with the pictures in \figref{lat-real}.

The resulting design can be compared with AGILE, a lighter instrument with a similar purpose and design but with the additional inclusion of a hard X-ray detector. Launched in April 2007, with about 1/10th of the LAT mass, AGILE was optimized for operation in the energy ranges 20--60~keV and 400~keV--30~GeV \citep{agile,agile10y}.

The Gamma-ray Burst Monitor (GBM) was selected as a secondary instrument on \fermi\ \citep{gbm}: comprising 12 sodium iodide (NaI) low-energy detectors and 2 bismuth germanate (BGO) high-energy detectors, it covers the entire sky not occulted by the Earth, in the energy range from $\sim 8$~keV to $\sim 20$~MeV; see \figref{lat-real}, right. The \fermi -GBM complements the capabilities of the LAT: together the 2 instruments are sensitive across more than 7 decades of energy, enabling the joint analysis of spectra and time histories of transient events, including GRBs and the electromagnetic counterparts of gravitational wave events \citep{grb-cat}.

\fermi\ was launched on June 11, 2008, and after the initial commissioning, configuration and calibration phase, the LAT began nominal science operations on August 4, 2008.

\subsection{The tracker (TKR)}\label{sec:tkr}

The LAT TKR is the game changer with respect to the predecessors. SSDs replace wire chambers, each detector giving the transverse coordinate of the crossing point of an iozining particle  with the readout pitch defining the spatial resolution. The ionizing energy is significantly lower for solid state detectors than for gas based detectors: $\sim 3$~eV for semiconductors versus $\sim 30$~eV for gas. As a consequence, more carriers are generated per ionizing particle, improving the signal-to-noise ratio. SSDs can be operated at a relatively low  voltage ($\sim 100$~V) with a small dark current, limiting the power consumption. The inherent radiation hardness of SSDs ensures this figure will not grow too much for the duration of the mission, even taking into account the expected radiation damage in space. On the other hand, SSDs of very large size are impractical: the TKR planes need to be assembled out of several smaller detectors. However, too much granularity is detrimental, due to the required support structures and readout electronics taking up space at the expense of active area, so an optimal compromise is necessary. In HEP trackers, multiple scattering in the Silicon volume can have a severe impact on the angular resolution of the detector and double-sided detectors are often advantageous, enabling the readout of both transverse coordinates in a single detector, thus limiting the mass at the expense of additional complexity and cost. This is not the case for the LAT TKR, where a large mass is required and, in fact, the mass budget is dominated by the conversion foils, so single-sided Silicon detectors are adequate. 

\begin{figure}[htbp]
  \centering
  \includegraphics[width=0.48\textwidth]{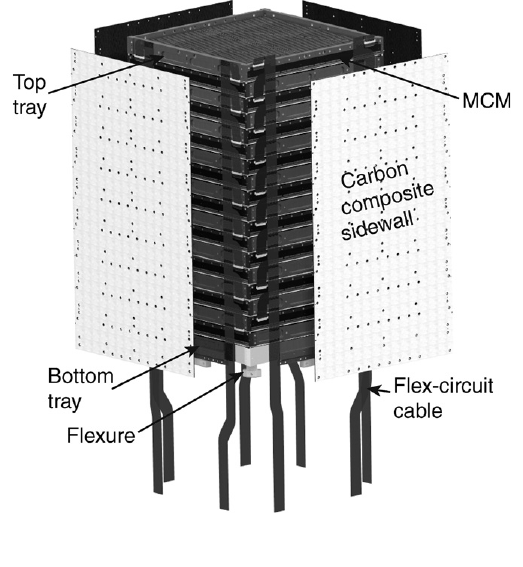}
  \includegraphics[width=0.48\textwidth,trim=0 -3cm 0 0]{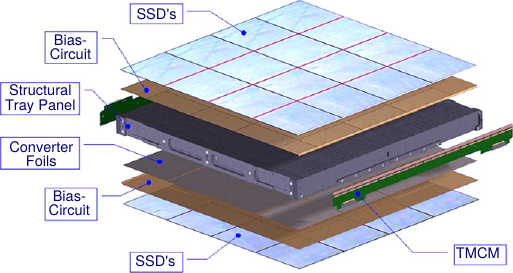}
  \caption{Left: rendition of one TKR tower, exploded, with stacked trays, readout electronics (``TMCM'') on the sides, and flex circuit cables extending below towards the tower electronics; from \citet{tkr-design-test} \copyright\ Elsevier. Right: exploded view of a TKR tray; from \citet{tkr-constr-test} \copyright\ Elsevier.}
  \label{fig:tracker}
\end{figure}

The TKR is divided into a matrix of $4\times 4$ identical towers; see \figref{tracker}, left. The basic TKR element is the \emph{tray}; see \figref{tracker}, right. 	  Individual SSDs are assembled longitudinally along the direction of the strips and micro-bonded in ladders of 4; then, 4 ladders are placed side by side to form a Silicon plane. Two SSD planes are located at the top and bottom of a tray, with the strips oriented in the same direction and the readout electronics placed on circuit boards (Multi-Chip Module, MCM) at 2 opposite edges. Tungsten (W) conversion foils are placed in between the active volumes, as close as possible to the bottom SSD plane. A TKR tower is assembled by stacking trays, each one rotated $90^\circ$ with respect to the previous one. This way the top plane of a tray and the bottom plane of the next one form an $x-y$ SSD layer, measuring both coordinates of a passing ionizing particle, and located immediately after the W conversion foil of the top tray. Stacking 19 trays creates 18 $x-y$ planes; the two SSD planes at the very top and bottom of a tower are not necessary and they are not included. Some parameters for the TKR are listed in \tabref{tkrsensor}. For more details on the design of the sensors see also \citet{tkr-ssd-design}.

As mentioned, most of the Coulomb scattering in the TKR occurs in the W foils, affecting the track development and reconstruction. To balance angular resolution against conversion efficiency, the TKR is divided into two sections. In the \emph{front} (or \emph{thin}) section, the 12 tracking layers are preceded by tungsten foils 0.1~mm thick (0.03 radiation lengths), giving relatively lower efficiency but better angular resolution. In the \emph{back} (or \emph{thick}) section, the first 4 tungsten foils are 6 times thicker, with opposite effects, and the 2 last layers have no conversion foils; we will discuss the reason for the missing last 2 foils when describing the trigger and data acquisition system. The  performance of the front and back sections in the TKR differ significantly and will be addressed separately in the section about instrument performance.

\begin{table}[htb]
  \centering
  \begin{tabular}{lcc}
    \hline
	SSD outer size					&	$8.95 \times 8.95$~\cmsq	\\
	Strip pitch						& 	228~\um					\\
	Floating strips					& 	no					\\
	SSD thickness					& 	400~\um					\\
	Depletion voltage				&	$<120$~V					\\
	Leakage current					&	$\sim 1$~nA/\cmsq\ at 150~V \\
	Breakdown voltage				&	$>175$~V					\\
	Fraction of bad channels		&	$\sim 0.01$\%				\\
	Number of SSDs tested			&	12500						\\
	Number of single strip tests	&	$\sim 30$M					\\
	Rejected SSDs					&	0.6\%						\\
    \hline
  \end{tabular}
  \caption{Some parameters for the TKR, including figures from the assembly and test. Adapted from \citet{tkr-perf-2-years}.}
  \label{tab:tkrsensor}
\end{table}

The choice of a readout pitch in SSDs is limited by manufacturing considerations and, in particular, by the development of the electric field in the detector. If needed, a fraction of the strips can be left floating, i.e. not connected to a readout, to properly bias the detector while limiting the number of channels. The readout pitch determines the hit spatial resolution, and therefore the angular resolution of the reconstructed tracks, in addition to the number of channels to be read.  At lower energies the resulting angular resolution is dominated by multiple Coulomb scattering, leading to an expected $1/E$ energy dependence, while at high energies one reaches the intrinsic resolution given by the strip pitch and lever arm. The chosen value for the SSD strip pitch of 228~\um\ enables the LAT to reach the intrinsic limit of $0.1^\circ$ for normal incidence at a few 10~GeV, close to the design energy upper limit of 100~GeV. For reference, decreasing the pitch by 40\% would improve the angular resolution at 10~GeV by 12\%, with the drawback of 24\% more channels in the TKR and consequently an even tighter energy budget per channel \citep{tkr-2003}. Due to this high spatial resolution, the internal alignment of the active elements and external alignment to the spacecraft are critical. We will mention the alignment of the TKR elements again when describing the calibration procedures.

%

The TKR readout electronics manage 885,000 channels, each one with a cumulative strip length of $\sim 35$~cm and strip capacitance $\sim 40$~pF. Given the number of channels and the power budget reserved for the TKR, the readout must employ $<300$~\uW/chn. The charge collected by each SSD strip enters an amplifier-shaper-comparator chain in one of the 64-channel front-end ASICs.  The amplified signal in a channel is discriminated by a single threshold and the pulse height is not measured, this limits the power consumption to $\sim 180$~\uW/chn \citep{tkr-perf-2-years}. Neighboring front ends are connected in a daisy-chain to instrument an entire TKR plane; digital controller ASICs at both ends control the front-end electronics and interface with the tower electronics. A simple trigger primitive is built from the logical OR operator of all the comparator outputs in a TKR plane, rising when at least one strip goes above threshold. The signal-to-noise ratio is excellent: for a minimum ionizing particle (MIP) crossing the SSD vertically, $\sim 5$~fC are deposited: for the nominal threshold value of 1.4~fC the hit efficiency is $>99.5\%$, while the noise occupancy is only $\sim 5\cdot 10^{-7}$ \citep{tkr-ro}. This electronic noise is one order of magnitude lower than the strip occupancy in orbit, caused by pulse tails of off-time cosmic-ray tracks. The shaping time is set to 1.5~\us : speed is not an issue, and in any case the scintillating detectors (CAL, ACD) impose an event timescale of the order of \us. Even so, when a MIP crosses a TKR plane the output remains high for $\sim 10$~\us\ due to details in the implementation of the baseline restoration circuit. Pile-up with the abundant background event can occur and is managed in the event analysis phase \citep{p7paper,pass8}. 

At the low end of the energy range, a \gammaRayHyph\ event deposits a significant fraction of its energy in the TKR ($\sim 50\%$ at 100~MeV). In addition, knowledge of the energy deposition profile in the TKR is useful for the event analysis (e.g. for the background rejection). With no pulse-height capabilities, a simple estimate of the energy deposited in one SSD is obtained by measuring the time the channel remains above the discriminator threshold, the  time resolution being defined by the internal 20~MHz clock. The $\sim 5$~fC deposited by a normal-incidence MIP correspond to a time-over-threshold (TOT) of $\sim 7.5$~\us; TOT is counted up to 50~\us\ (6 MIPs) to limit the readout delays, while linearity was shown to be good well beyond this limit \citep{tkr-design}.

The TKR front-end ASICs include a charge injection system, for calibration purposes: a capacitor is connected to each input and a step voltage, generated by a digital-to-analog converter, can be applied to selected channels \citep{tkr-ro}.

\begin{figure}[htbp]
  \centering
  \includegraphics[width=0.48\textwidth,trim=0 -1cm 0 0]{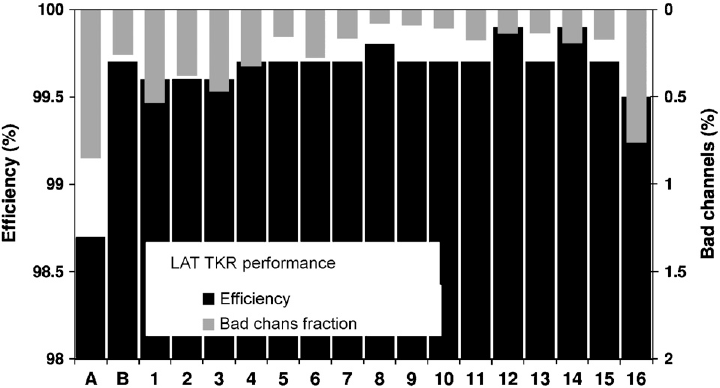}
  \includegraphics[width=0.48\textwidth]{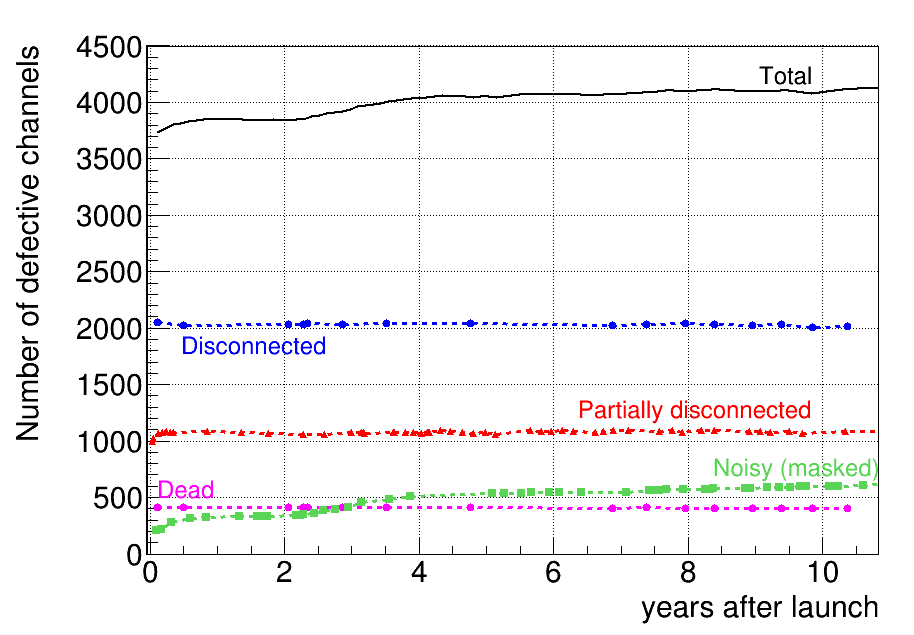}
  \caption{Left: Hit efficiency and fraction of dead channels in the 18 TKR modules built; from \citet{tkr-calib} \copyright\ Elsevier. Right: Dead and noisy channels in the TKR as observed in orbit; from \citet{instr10y} \copyright\ AAS.}
  \label{fig:strips}
\end{figure}

When a \gammaRay\ converts inside the TKR, it is usually inside a tungsten foil: each x-y SSD layer contributes only $\sim 0.01$~\rl. If the secondaries are detected in the very first silicon layers the uncertainty on the conversion location caused by Coulomb scattering and lever arm is $\sim 10$~\um, below the intrinsic resolution of the SSDs, but this degrades by more than an order of magnitude if the hits in the first layers are missed, leading to a loss in angular resolution of up to a factor $\sim 2$, at 100 MeV. It is therefore critical to maintain a very high hit efficiency. In particular, \gammaRayHyph\ trigger and hit efficiencies would be severely affected if a sizable number of the channels in the TKR were not functional. In \figref{strips}, left, MIP detection efficiency and fraction of bad channels are shown for the 18 TKR tower modules that were produced. Module A was the first ever built, the experience gained in the process is evident by the improvement in the following modules; despite Module A being slightly worse than the rest of the modules, it is still within the requirements (MIP efficiency $>98\%$), so it qualified for inclusion in the LAT. Module 16 was built with non-flight modules, and together with Module 8 was put aside as a spare part and later included in a so-called ``calibration unit'', discussed later.

In \figref{strips}, right, the evolution in the number of bad channels is shown as a function of time elapsed during the mission. After 10 years of operation in orbit, the number of bad strips amount to 4087 (0.46\%), a small increase with respect to the 3661 bad channels at launch. Bad channels are broken down into several categories: \emph{dead} channels appear to have a dead preamplifier, showing no signal and zero noise; \emph{disconnected} channels give no signal and very low noise, compatible with a floating input not connected to the SSD; \emph{partially disconnected} strips have intermediate noise levels, indicating a broken connection somewhere along the ladders; \emph{noisy} channels are due to some unspecified problem and have to be masked to prevent them from generating trigger requests and data hits. Remarkably, while the number of dead channels increases slowly with time, the number of dead and disconnected channels diminished slightly, indicating some unknown kind of reversible damage \citep{instr10y}.

\subsection{The calorimeter (CAL)}\label{sec:cal}

Inorganic scintillators, and Thallium-doped CsI in particular, are well suited for the construction of a large, segmented calorimeter \citep{knoll}. CsI(Tl) is very bright (54 photons/keV) but relatively slow (decay time of about 1~\us\ for \gammaRays), so it is well suited to applications where the particle rates are not too high. The maximum of the light emission occurs at around 550~nm, well-suited for Silicon photodiode readout. Being a relatively low-hygroscopic material, it makes it unnecessary to seal with passive materials of low EM stopping power. It is quite robust, with no cleavage planes and reasonably radiation hard, and therefore it is widely used for space applications.

The CAL, in order to be placed below the TKR must have the same modularity (16 tower modules) and lateral dimensions, while the vertical dimension, or thickness, is defined by balancing three goals: wide energy range (the thicker the better), wide FOV (squat aspect ratio), and acceptable total mass. With the goal of pushing the energy range above 100~GeV the optimal value thickness was set 8.6 \rl\ of CsI, or 16~cm. Part of the EM showers will escape from the bottom, the sides and along gaps in the modular structure, especially at high energies. To correct for this lost fraction, the shower development must be reconstructed. Several solutions to improve the imaging power were investigated (scintillating fibers, sampling calorimeter, pre-shower) and were discarded due to the small improvement in performance, often accompanied by a loss in energy resolution at the low end of the design energy range \citep{cal-letter}.

\begin{table}[htb]
  \centering
  \begin{tabular}{lcc}
    \hline
    	Total mass							&   1376 kg \\
   Scintillator material						&	CsI(Tl)							\\
	Crystal dimensions						&	$26.7$~mm$\times 20$~mm$\times 326$~mm	\\
	Crystal mechanical tolerance			& 	0.3~mm					\\
	Wrapping					    &	aluminized Mylar \\
	Number of crystals				&	1536					\\	
	Electronics channels			&	6144	\\ 
	Readout dynamic range			&  	$\sim 5\cdot10^5$	\\
	Required longitudinal position  &	$<1.5$~cm	\\ 
	\ resolution (1$\sigma$)		&  \\
	Required energy resolution 	    &  $<20\%$ ($<100$~MeV, CAL only) \\
	\ (on axis, $1\sigma$)			&  $<10\%$ (100~MeV--10~GeV) \\
									&  $<20\%$ (10--300~GeV)\\
    \hline
  \end{tabular}
  \caption{Some parameters for the CAL, including figures from the assembly and test. Compiled from \citet{latpaper} and \citet{cal-thesis} .}
  \label{tab:calsensor}
\end{table}

\begin{figure}[htbp]
  \centering
  \includegraphics[width=0.48\textwidth,trim=0 0 0 0]{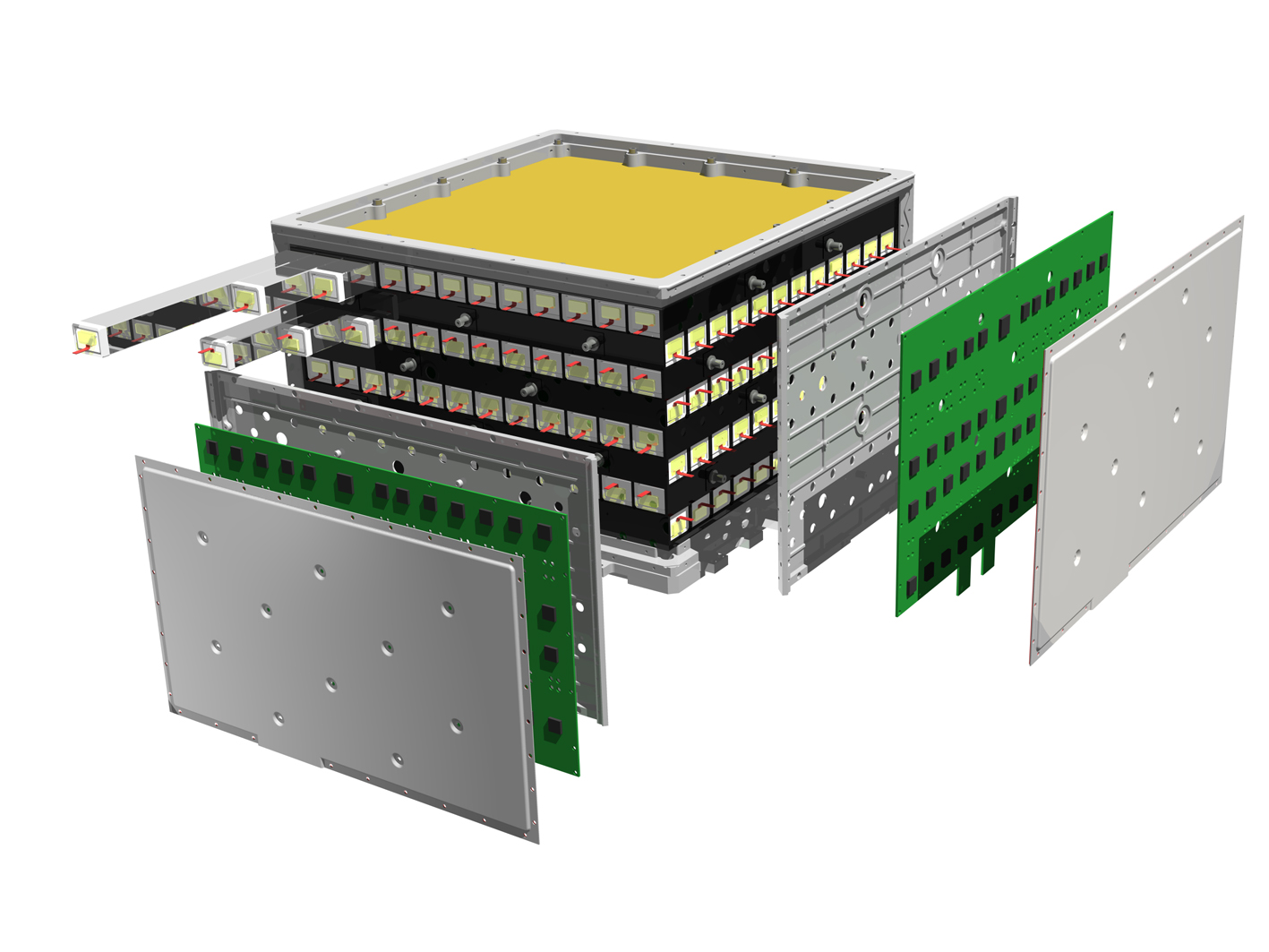}
  \includegraphics[width=0.48\textwidth]{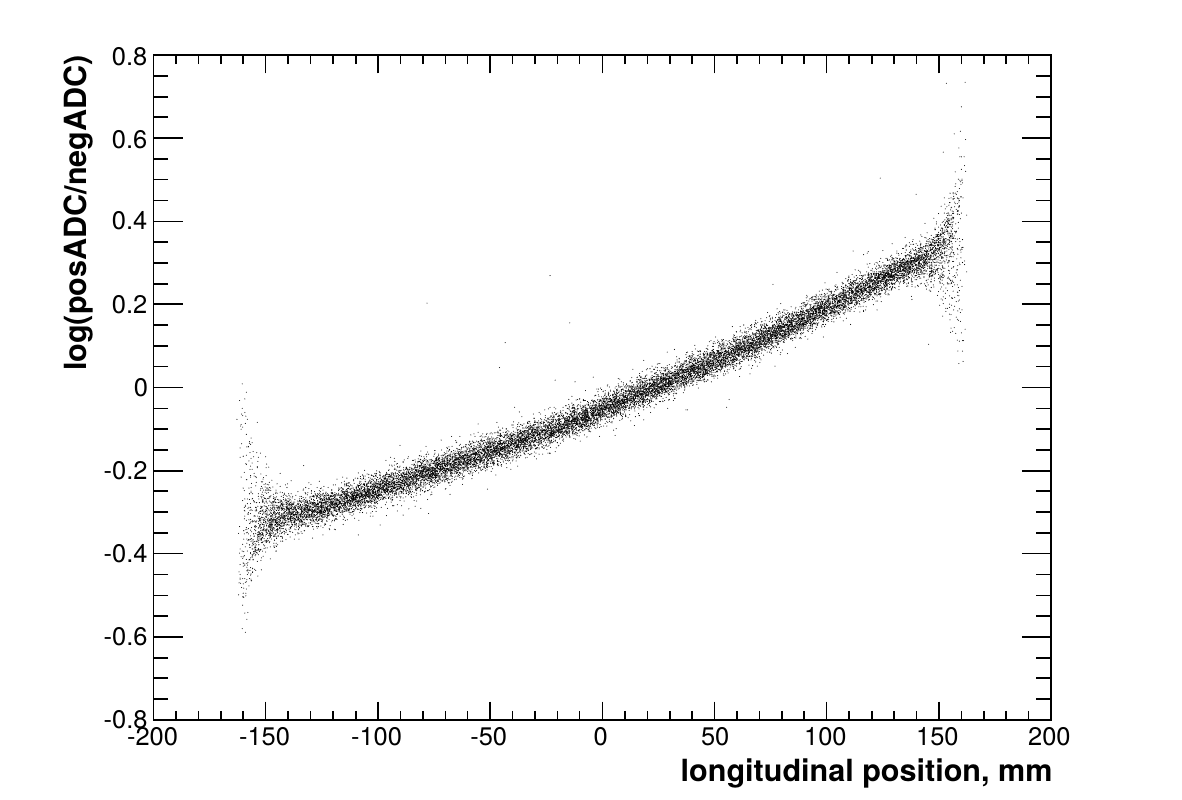}
  \caption{Left: exploded view of a calorimeter module. Right: asymmetry in light collection at both ends of a crystal versus position for normally incident muons. From \citet{latpaper} \copyright\ AAS.}
  \label{fig:cal}
\end{figure}

The CAL is divided into 16 identical modules, one per tower; some parameters and requirements are listed in \tabref{calsensor}. In each module, there are 8 layers of 12 parallel crystals in hodoscopic arrangement, each layer rotated $90^\circ$ with respect to the previous one.  Each crystal is individually wrapped in reflective foil, and two photodiodes are glued at each end: a large one (147~mm$^2$ in area, $2$~MeV to $1.6$~GeV in energy range) and a small one (area 25~mm$^2$, energy range $100$~MeV --$70$~GeV). See \figref{cal}, left, for an artist's rendition of the structure.

The lateral dimensions of the crystals make it possible to sample the shower development: lateral size and thickness are close to the Moli\`ere radius (3.53~cm) and the radiation length of CsI (1.86~cm). The position along the longitudinal direction is obtained from the asymmetry in light collection at the two ends, caused by attenuation inside the crystal. The asymmetry, and hence the spatial resolution, is optimized by controlling the surface treatment of the longitudinal surfaces: polishing improves the yield but impairs the spatial resolution, while roughening (e.g. lightly scratching with an abrasive material) causes the opposite effects. The longitudinal position resolution in the CAL depends on the deposited energy, ranging from a few mm at 1~MeV of deposited energy to less than 1~mm at $>10$~GeV \citep{cal-thesis}. In \figref{cal}, right, the logarithm of the asymmetry in light collection is plotted against longitudinal position for normally incident muons (deposited energy $\sim 11$~MeV). 

More than 2000 crystals were procured from the manufacturer with quality slowly improving as the manufacturing process was refined. Overall, $\sim 80\%$ passed all the mechanical and optical requirements without changes, and most of the rest were barely outside specifications and were adjusted with an additional surface treatment during the testing phase.

The dose the crystals were expected to receive was estimated before launch to be $\sim 40$~Gy per 10 years of mission. Radiation hardness was evaluated in the acceptance tests: crystal samples for each production batch were irradiated with $100$~Gy of $1$-MeV \gammaRays, leading to an average loss of light yield of $12\%$ (maximum loss $27\%$), well within the requirement of $<50\%$ \citep{cal-rad}. Accounting for the 2.5 safety factor, the average loss is in good agreement with the observed $\sim 6\%$ yield loss in 10 years, see \figref{calevol}.

The readout electronics, placed on the sides of the CAL modules, must fulfill the demanding task of operating across a wide dynamic range with low power consumption and minimal dead time. The overall scheme is similar to that of the TKR: ASIC readout electronics, a front-end analog chip with an amplifier-shaper-comparator chain (shaping time $\sim 3.5$~\us) for each crystal end, and separate digital readout controller ASICs. The output of each diode is split into two track-and-hold circuits with gains differing nominally by a factor of 8. This enables the coverage of the large dynamic range with 4 readout ranges chosen automatically, in the ratios 1:8:64:512. In the front-end an additional fast shaping amplifier ($\sim 0.5$~\us) is included for trigger discrimination. Two threshold discriminators at each crystal end, one per photodiode, generate two trigger requests indicating a low- or high-energy deposition; nominal settings are 100~MeV and 1~GeV, respectively. See \citet{latpaper,cal-design} for more details. To decrease the readout deadtime and the data volume, an ``accept'' threshold is set for each crystal end (nominally 2~MeV): signals below threshold are suppressed; dead time per event is less than $20$~\us.

Similarly to the TKR front-end electronics, a charge injection system is implemented to calibrate the input channels individually. In addition, the significant overlap between the readout energy ranges makes it possible to cross-calibrate the channels, and simultaneous readout of the four ranges is available and used in calibration runs.

Twenty CAL modules were assembled, of which 16 were integrated into the LAT. Three additional modules are used in the so-called ``calibration unit'', while one engineering module, not completely identical to the others, was used in the beam-tests, which are discussed in the calibration section.

\subsection{The anti-coincidence detector (ACD)}\label{sec:acd}

The lack of a time-of-flight detector on board to tag unwanted upward-moving particles makes the performance of the anti-coincidence detector critical for the success of the instrument \citep{acd-design}. In the \fermi\ orbit, charged particles outnumber \gammaRays\ by five orders of magnitude. Under these conditions, the rate of trigger requests from the TKR alone averages to several kHz, and most of these trigger requests should be rejected in order to limit the dead time to a reasonable figure. Even so, the triggered events are mostly background that must be discarded on board to bring the bit rate within the available downlink bandwidth. As mentioned, the presence of the CAL complicates the matter: in high-energy showers, secondaries will escape and reach the lower parts of the ACD, potentially causing a veto (calorimeter backsplash). In fact, this caused a reduction of the high-energy efficiency for EGRET, by a factor $\sim 2$ at 10~GeV, with respect to the efficiency at 1~GeV \citep{egret-calib}.

\begin{figure}[htbp]
  \centering
  \includegraphics[scale=1]{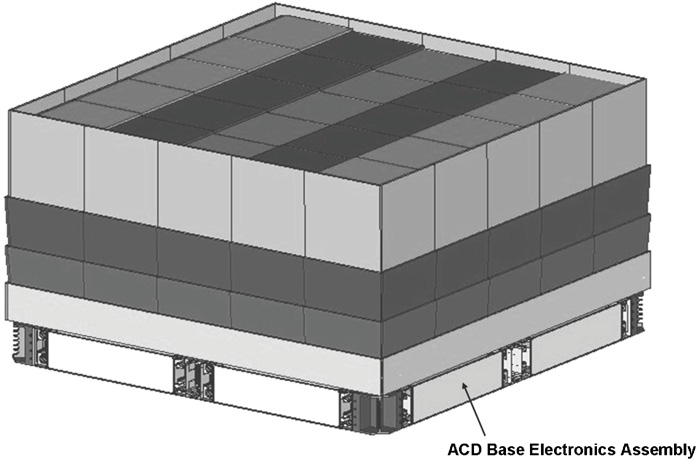}
  \caption{ACD structure, from \citet{latpaper} \copyright\ AAS.}
  \label{fig:acd}
\end{figure}

Plastic scintillators are well suited for an anti-coincidence detector: they are sturdy, inexpensive, can be machined into complex shapes and easily cover a large surface, and can achieve a hit efficiency greater than 0.999 \citep{knoll}. Embedding wavelength-shifting (WLS) fibers in the material enables the light signal to be brought to the optical readout, which can be conveniently moved away from the FOV. 

The ACD is a square hat covering the top and sides of the LAT, extending down to cover the entire TKR, with a total surface of $8.3$~m$^2$ \citep{acd-scint-wsf}. It is segmented into 89 tiles of various shapes and sizes with areas ranging from $\sim 450$ to $\sim 2500$~\cmsq\ and 1~cm thick (the five tiles in the top middle row are 1.2~cm thick, to compensate for the greater distance from the readout with a slightly larger signal), see \figref{acd}. Each tile is a  polyvinyl toluene plastic scintillator with 64 grooves machined uniformly on the surface, where 1~mm diameter WLS fibers are embedded. The fibers from a tile are split into two symmetric bundles and run to 2 photomultiplier tubes (PMT) for readout. Light yield uniformity is typically $>95\%$ across most of the tile surface, except for the 2~cm-wide region around the borders, where it remains $>75\%$.  Finally, each tile is individually wrapped with light-reflective foils and then a black light-tight envelope.
In addition to the plastic tiles, cable-like ribbons of scintillating fibers are used to improve the seal along directions where tile overlap was not possible; the ribbons themselves have detection efficiency $>90\%$. 

The modular structure improves the robustness of the whole system: a puncture in the ACD would disable one tile only, with a limited impact on the overall performance. In addition, segmentation allows for a significant improvement over the monolithic predecessors: the reconstructed direction of the primary and the shower development can be correlated to the location of the hits in the ACD, and the simple veto condition ``hit in the ACD'' can be expanded to mitigate the calorimeter backsplash. A possible disadvantage is leakage through the borders of the tiles where the hit efficiency is lower. This is alleviated in two ways: misalignment and overlap. The ACD is built with a 5x5 structure, so the edges do not match the gaps between the towers where the tracking uncertainty is highest, helping with the correlation of tracks in the TKR and hits in the ACD. In addition, the top tiles are overlapping along one direction and bent tiles are used for the top edges. The remaining gaps are sealed from the inside with the ACD ribbons. Notably, the lowest tiles, closest to the CAL and outside the design FOV, are not segmented: there is no need to try and recover \gammaRayHyph\  events passing through these. 


The front-end electronics are located on two opposite sides at the base of the LAT, divided into 12 circuit boards, each managing 18 channels. The electronics assemblies contain the redundant HV power supplies, the analog front end ASICs and the digital readout controllers. The ACD readout generates the fast trigger signals and a sample-and-hold signal for pulse-height analysis (PHA).
Setting the veto threshold defines the efficiency for charged particle detection, with a requirement of $>0.9997$ for a MIP. In the ACD design the tile readout has two thresholds: on board a threshold of $\sim 0.45$~MIP is used for the initial rejection of charged particles, and on ground the analysis threshold of $\sim 0.3$~ MIP for the final analysis brings the detection efficiency to $\sim 0.9999$.

\begin{figure}[htbp]
  \centering
  \includegraphics[width=0.8\textwidth]{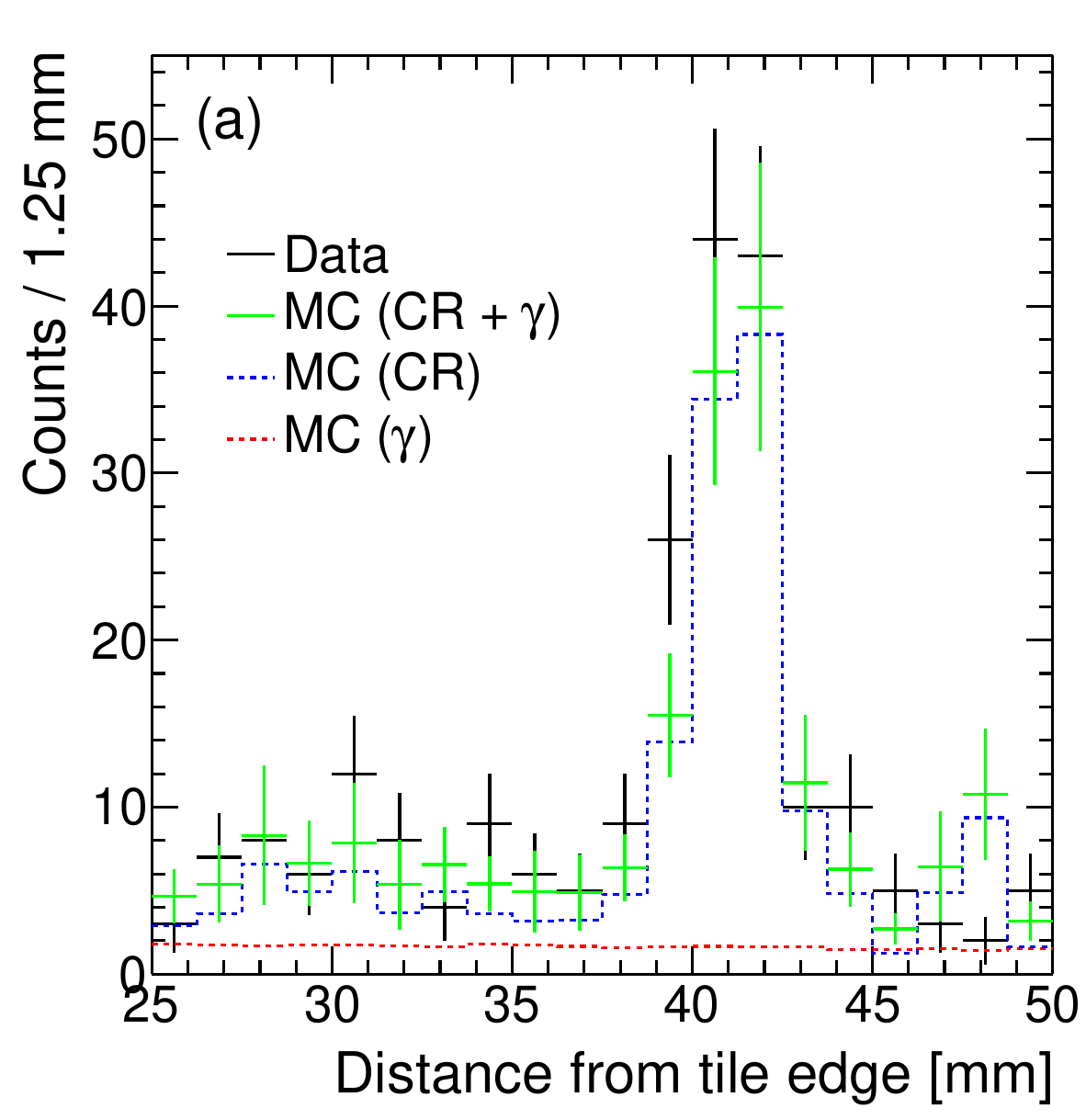}
  \caption{Reconstructed direction of residual background events: minimal distance of the extrapolated intersection point from the edge of the tile for real data and simulated events. From \citet{p7paper} \copyright\ AAS.}
  \label{fig:acd-bkg}
\end{figure}

To appreciate the overall efficiency of the ACD, we can consider the background events remaining after the background rejection is performed on the ground. Detection efficiency is not perfectly uniform over the ACD, causing some regions with relatively smaller efficiency to act as a path in for background events. In \figref{acd-bkg}, the minimal distance of the intersection point of the extrapolated TKR track from the edge of the traversed ACD tile is shown for a sample of real and simulated events, after the level of background rejection recommended for the observation of point sources is applied. An excess can be seen at $\sim 40$~mm from the edge, corresponding to the location of many mounting holes; a significant number of background particles remaining in the dataset are passing through those. This region of the phase space can be eliminated in the event analysis process at the cost of a small loss in effective area, and a tighter background rejection can thus be achieved. 
Other background populations are visible in the Monte Carlo simulations but hard or impossible to address in reality, e.g. anything producing secondary \gammaRays\ outside the ACD. As an example, protons can undergo inelastic scattering in the passive materials surrounding the ACD, producing low-energy \gammaRays; a clear association of the events with the surroundings of the ACD is problematic, not least because of  the relatively large uncertainties in the reconstructed direction at low energy.

\subsection{Data acquisition and event analysis}\label{sec:daq}

LAT data taking is organized in \emph{runs}, each usually spanning one orbit or the time between exiting and entering the South Atlantic Anomaly (SAA). The signals from the subsystems must be collected, processed and sent to the ground for further analysis. This process can be divided into two phases: a hardware one in which channels are latched and read, and data is collected by the LAT Event Processing Unit (EPU). In the second phase, a software one,  the data are processed by the EPU and stored in the Solid State Recorder (SSR), ready for downlink.

At the core of data collection is the \emph{event trigger}, which ultimately determines the dead time of the instrument. The LAT is, by design, a \gammaRayHyph\ detector, so the the main role of the trigger is to activate on \mygamma -like events. On the other hand, other kinds of events are necessary in order to calibrate and monitor the subsystems (e.g. MIPs and heavy ions). The LAT operates with a flexible trigger system, where different \textit{trigger engines} run at the same time, based on several \textit{trigger primitives} (or \textit{trigger requests}). Most trigger primitives are generated by the subsystems (TKR, CAL, ACD) when a suitable energy deposition occurs in the active volumes of the detectors; a few are generated internally by the DAQ system. 

\begin{itemize}
\item \techname{TKR} is issued when three consecutive x-y layers in the tracker have a signal above threshold (nominally $0.25$~MIP), indicating a possible particle track. 
\item \techname{CAL\_LO} is issued when a calorimeter crystal has a signal above the low-energy threshold (nominally 100~MeV).
\item \techname{CAL\_HI} is issued when a calorimeter crystal has a signal above the high-energy threshold (nominally 1~GeV).
\item \techname{VETO} is issued when an anticoincidence tile has a signal above the low-energy threshold (nominally 0.3~MIP). 
\item \techname{ROI} is issued when a \techname{TKR} primitive happens in coincidence with a \techname{VETO}: each tower has a list of associated anticoincidence tiles to check for coincidence.
\item \techname{CNO} is issued when an anticoincidence tile has a signal above the high-energy threshold (nominally 25~MIP). 
\item \techname{PERIODIC} is the only special primitive affecting normal operation. It is issued at a constant frequency (nominally 2~Hz).
\end{itemize}

The ``three'' in the definition of the \techname{TKR} primitive is one of the few non-configurable numerical parameters in an extremely versatile system. We also note that \gammaRays\ converting just before either of the two last TKR layers would not cause a trigger request, which is the reason why there are no tungsten foils in the bottom two trays.

\begin{figure}[htbp]
  \centering
  \includegraphics[width=0.66\textwidth]{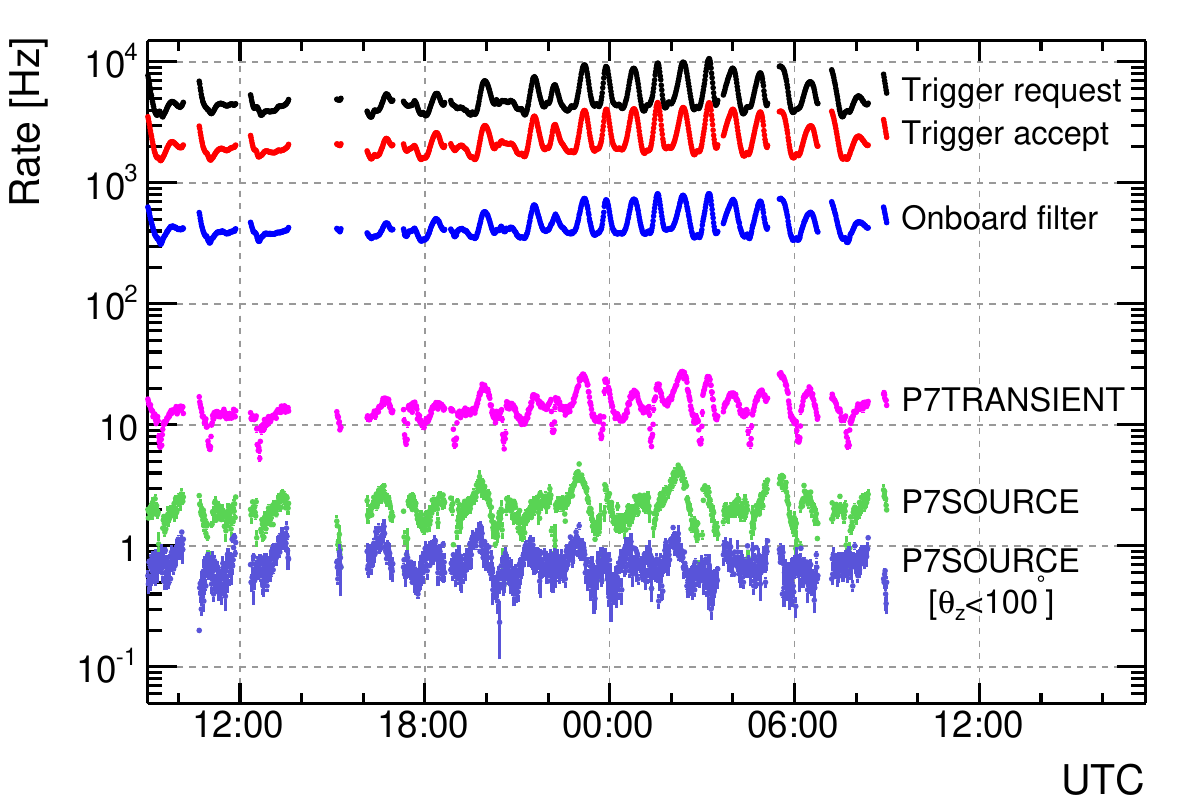}
  \caption{Rates at several stages of the data acquisition process during a typical day of standard operation. Rates from top to bottom: trigger requests from the subsystems, issued LAT triggers, output of the
on-board filter, after a loose \gammaRayHyph\ selection, after a
tighter selection, and including an additional cut on the \mygamma\ zenith angle to reduce contamination from the Earth limb. From \citet{p7paper} \copyright\ AAS.}
  \label{fig:rates}
\end{figure}

Trigger engines are built with the above primitives, with the relevant ones during normal operation in orbit listed in \tabref{triggers}. To limit bandwidth usage, some are \emph{prescaled} by a factor $n$, i.e. only 1 trigger request in $n$ is acknowledged. The basic \gammaRayHyph\ trigger is number 7, corresponding to a track candidate in the TKR, no veto from the ACD, no large energy deposit in the CAL, and no prescaling. The resulting cumulative trigger rate averages $\sim 1.5$~kHz, see \figref{rates}.

\begin{table}[ht]
  \begin{center}
    \begin{tabular}{cccccccrr}
      Engine & \techname{PERIODIC}& \techname{CAL\_HI}& \techname{CAL\_LO}&  \techname{TKR}& \techname{ROI} &  \techname{CNO} & Avg. rate [Hz] &  \\
      \hline
      3 & \isreq &        &        &        &        &   & 2 & \\
      4 & \isexc &        & \isreq & \isreq & \isreq & \isreq  &  200 & \\
      5 & \isexc &        &        &        &        & \isreq  &  5 & $\dagger$ \\
      6 & \isexc & \isreq &        &        &        & \isexc  &  100 &  \\
      7 & \isexc & \isexc &        & \isreq & \isexc & \isexc  & 1500  & \\
      8 & \isexc & \isexc & \isreq & \isexc & \isexc & \isexc  & 400 & $\star$\\
      9 & \isexc & \isexc & \isreq & \isreq & \isreq & \isexc  & 700 &  \\
     10 & \isexc & \isexc & \isexc & \isreq & \isreq & \isexc  & 100 & $\ddagger$ \\
    \end{tabular}
    \caption{Definition of the standard trigger engines: primitives used
      (\isreq: required, \isexc: excluded) and
      average rates; adapted from \citet{p7paper}.
      Trigger engines 0 to 2 are not relevant for science operation. $\dagger$: prescaled by 250. $\ddagger$: prescaled by 50. $\star$: currently disabled.}
    \label{tab:triggers}
  \end{center}
\end{table}

The next step is the \emph{event filter}, fulfilling the role of reducing the event rate until the data stream is compatible with the available downlink bandwidth. This system is also configurable, allowing several independent filters to run at the same time; three filters are active during nominal operation.

\begin{itemize}
\item \techname{GAMMA} is designed to accept \gammaRayHyph\ candidates;
\item \techname{MIP} is designed to select heavy-ion candidates;
\item \techname{DIAGNOSTIC} accepts all events taken by the \techname{PERIODIC} trigger and an unbiased sample of all other trigger engines, prescaled by a set factor (nominally 250).
\end{itemize} 

Most of the \gammaRay\ events in the science datasets passed the \techname{GAMMA} filter: a sequence of conditions are evaluated for each event, in order of complexity. This includes a rudimentary track reconstruction in the later stages. The average event rate, after this filter, is reduced to an average $\sim 400$~Hz, see \figref{rates}. This translates to an average of 1.5 Mbps sent to the SSR.

Each event is timestamped and the detector dead time is recorded, with the intrinsic time resolution of 50~ns set by the LAT internal clock, operating at 20~MHz. Timestamping relies on the absolute time provided by the internal GPS receiver, plus a precision 20-MHz scaler synchronized to the GPS Pulse-Per-Second signal. Tests before launch indicated that LAT GPS times are maintained within 20~ns of UTC time, and event timestamps are accurate within 1~\us\ \citep{onorbitcalib}. Timing is monitored during the mission with an accuracy of a few $\mu$s by measuring the period of millisecond pulsars \citep{instr10y}. Instrumental dead time is dominated by the time required to latch the front ends and read out the event, with a minimum of $26.5$~\us\ per event. Dead time is, on average, $\sim 8$\% outside the SAA, with small variations correlated to the trigger rate \citep{onorbitcalib}. The value is downlinked with the event data, since it must be accounted for in order to calculate the fluxes of \gammaRayHyph\ sources. Additional dead time due to data loss (on board or on ground) is well below $1$\%.

In addition to the aforementioned nominal science operation mode, the LAT can be operated in \emph{dedicated mode}: in this mode the detector electronics and the trigger system are configured to acquire data for calibration and synchronization purposes. While in dedicated mode the \techname{PERIODIC} trigger can be put in charge-injection mode, instructing the front end electronics to feed charge pulses directly to the preamplifiers, making it possible to calibrate all readout channels individually. Since this mode is incompatible with science data taking, dedicated mode must be periodically enabled during \emph{calibration runs}.

The LAT transmits to the ground an average of 16~GB of compressed data every day. Data are processed at the computer farm at the SLAC National Accelerator Laboratory (SLAC): over 3000 CPU cores are available to promptly reconstruct and analyze the event data and publicly release them as soon as possible. Another 1500 CPU cores are available at the IN2P3/CNRS facilities in Lyon, France, and are commonly used for Monte Carlo simulations.

Calibration datasets undergo a separate, dedicated analysis. In addition, the abundant background events are the target of dedicated analyses to evaluate the fluxes of cosmic-ray electrons, positrons, and protons \citep{electrons,elepos,eleanyso,protons}. From here onward, we focus on the \gammaRays, examining in some detail the process on the ground that leads to the creation of the \gammaRayHyph\ dataset.

\emph{Reconstruction} is the process of translating calibrated data from the detectors (deposited energy, hit locations, \dots) into a description of the physics behind them (tracks, energy of particles, volumes crossed, \dots). \emph{Event analysis} is the assembly of such information into a \gammaRayHyph\ event, with an incoming direction, energy, time, and ancillary quantities (quality of the energy and direction estimates, probability of being a background event, \dots). Finally, a set of \emph{cuts} on the available variables defines an \emph{event class} and, in practice, an event set. 
Releasing an improved event processing procedure is a major task: in addition to the effort in designing and validating the new algorithm, after deploying the new version to the real-time data analysis pipeline, all the past data in the archives must be reprocessed. From the beginning of the nominal operations in August 2008 to November 2013, the \Psix\ reconstruction and event analysis scheme was in place, developed prior to launch. From November 2013 to June 2015 \Pseven\ was employed, featuring the same reconstruction as the predecessor but a significantly improved event analysis scheme. The reconstruction and analysis procedure, and the several validation procedures that are performed on LAT data are described in great detail in \citet{p7paper,latpaper}. Since June 2015, \Peight\ is operational, featuring novel reconstruction algorithms and a new event analysis \citep{pass8}; the version currently in use is the third release, featuring a slightly improved background rejection \citep{p8v3}. Thus, in order to identify an event set, one must name the reconstruction and analysis procedure (e.g. \techname{P8R3} for \Peight\ release \techname{3}) and the selection cuts, usually described with a byname suggestive of the intended use or strictness of the background rejection (e.g. \techname{SOURCE} for the analysis of point sources). 

The instrument performance differs for each event class, with more stringent cuts decreasing efficiency and generally improving the quality, e.g. in terms of resolution, residual background level, etc; see the description of the LAT performance in the dedicated section. In all data releases, photon events have been partitioned into two conversion types (\emph{front} and \emph{back}), given the significant difference in performance, see \figref{perfplots}. Since \Peight, this has been expanded and generalized into the concept of event types,  event subsets for which performance is evaluated and provided, and which can be included or excluded from the scientific analysis. On top of the conversion type partition, two new event type partitions are available: \emph{psf} event types, indicating increasing quality of the reconstructed direction, and \emph{edisp} event types, indicating increasing quality of the energy reconstruction \citep{p8data-site}. If the angular and/or energy resolution are critical, one can select only the better event types for the analysis, at the price of a smaller effective area.

\subsection{Operation}\label{sec:operation}

The latitude of the launch 
site, at Cape Canaveral Air Force Station Space Launch Complex 17-B, set the initial orbit inclination at $28.5^\circ$. Once a circular orbit at $550$~km above sea level was reached, the remaining fuel was used to reduce the orbit inclination  to $25.6^\circ$, thus reducing the time spent in the SAA. In this region, HV power supplies in the ACD are turned off to protect the PMTs, thus regular data taking is disabled. For the operation of the LAT, the SAA is defined by a 12-vertex polygon, stop and start commands are issued 30 seconds before entry and after exit. Notably, the SAA polygons for the LAT and GBM differ, see e.g. the case of GRB170817A \citep{grb17a-gbm,grb17a-lat}: at the time of the GRB the GBM was outside its SAA polygon and taking data but the LAT was already inoperative. 

The orbit of \fermi\ is slowly decaying, with an altitude loss more pronounced during the solar maximum: in 10 years, the altitude decreased by about 20~km, with a corresponding change in orbital period from 95.7 to 95.3~minutes. 

The attitude profile is determined in order to maximize the uniformity of exposure across the sky, leveraging the wide FOV of 2.4~sr. The Earth is a very bright source of \gammaRays\  \citep{limb09} and can cause excessive dead time saturating the trigger system, so it is best kept outside the FOV. In the nominal observation profile, in survey mode (``scanning''), the spacecraft rocks north and south of the orbital plane on alternate orbits. As a consequence, the LAT boresight is offset from the zenith
toward either the north or south orbital poles by a characteristic
rocking angle. Initially, the rocking angle was set at $35^\circ$. This value was later changed a few times to optimize the operation and performance, to finally set at $50^\circ$. After the anomaly of March 2018, discussed later, the pointing strategy is more complex due to operational constraints. In addition to the uniformity of exposure across the sky, the rocking angle affects the average temperature of the spacecraft batteries: the larger the rocking angle, the more time is spent tilted away from the relatively warmer Earth, thus improving the cooling and increasing the mission lifetime. Within a given orbit, \fermi\ also executes a slow roll about the boresight to maintain an optimal orientation of the solar panels with respect to the Sun.

\fermi\ can be set in pointed observation mode, so that the LAT points in the direction of a target. This can be requested from ground (e.g. for Target Of Opportunity observations, transients, etc.) or initiated autonomously (e.g. when the onboard analysis detects a GRB candidate satisfying given requirements). In pointing mode, the target is kept close to the boresight, but when the Earth limb approaches the FOV, \fermi\ rotates to maintain a fixed Earth avoidance angle (nominally set to $20^\circ$) between LAT axis and Earth limb. In case an occultation occurs, the telescope switches to a roll along the Earth limb at a set angle (nominally $50^\circ$) with the appropriate speed to catch the target as it rises from occultation. 

Calibration runs are  performed routinely to evaluate and monitor the instrument performance. These require running the instrument with a special configuration and/or with a specific attitude profile.


Maintaining an optimal temperature is critical for the survival of the instrument and for ensuring the quality of data: thermal radiators and active heater elements operate to keep all subsystems within the allowed temperature range. 
If a major problem arises, the LAT is automatically powered off by the \fermi\ spacecraft. The temperature is maintained in a survival temperature range by the survival heaters on board, controlled by the spacecraft: the LAT can survive for an indefinite period of time in this state, while the Fermi Mission Operations Center at NASA's Goddard Space Flight Center and the LAT experts within the LAT Collaboration act to solve the problem.

On \mydate{2008}{July}{31}, an intermittent short in the wiring caused several temperature readouts to falsely appear too cold, outside the safe range, causing a power-off. The affected alarms have been disabled; in any case, the large thermal mass of the LAT allows ground operators to see any real temperature changes and react before the temperatures change too much.
On \mydate{2009}{March}{11}, a software error occurred in the LAT computer, causing a chain of other errors that ultimately caused the spacecraft to go into its safe-mode and powering off the LAT. The LAT was restarted by ground commanding and the errors in the computers were identified in the diagnostic data and fixed in a subsequent flight software update.

On \mydate{2018}{March}{16}, \fermi\ went into safe-mode and powered off the LAT, because the -Y solar panel stopped moving. The LAT remained powered off for
over 17 days, as the solar panel problem was investigated. 
Power up proceeded without problems and on April 2, science data taking resumed. Due to the large thermal inertia of the CAL and its temperature-dependent performance, the data were flagged as unfit for science analysis for a few days, until nominal operation was resumed 23 days after the shutdown, the longest interruption since launch. 
The affected solar panel on Fermi remains stuck since March 2018, and the rocking profile for all-sky survey was replaced with periods of various alternating rocking angles, keeping the \fermi\ power system operating nominally with minimal changes to the LAT sky exposure. Autonomous repointing 
has also been disabled.

Other rare issues require direct intervention by the Fermi Mission Operations Center. On \mydate{2012}{April}{03} the \fermi\ thrusters were fired briefly to avoid a predicted close approach with another satellite, resulting in a minimal impact on the orbital parameters. A similar maneuver was considered for a few days in April 2010 for another close approach and was canceled once the probability of a collision dropped down to acceptable levels (the orbit predictions become more accurate the closer in time they are to the conjunction).

The performance of the LAT subsystems and of the ground infrastructure are constantly monitored through the LAT Data Quality Monitor (DQM) system \citep{instr10y}. During the data processing on the ground, histograms are generated and made available to the scientists on duty, while automated alarms are issued if a quantity deviates from the allowed range. Monitored quantities include temperatures, pedestals and gain of the detectors, channel rates, and event rates. About 12,000 parameters are monitored and the DQM system makes about 4100 checks on parameter ranges. Quantities that vary significantly during an orbit due to dependence on the attitude or to the geomagnetic coordinates are parameterized as a function of the relevant variables, leading to normalized quantities that are easier to monitor. If the DQM system identifies a problem that can potentially influence the data quality, a bad time interval (BTI) is marked in the data file. As of the end of 2021 the cumulative bad time amounted to less than 10 days, in large part due to the aforementioned shutdowns and the following temperature stabilizing periods after restart ($\sim 70\%$); almost all the remaining BTIs are caused by solar flare activity. During solar flares, X rays hitting the ACD can cause excessive veto signals, reducing the sensitivity to \gammaRays. Under these conditions the LAT performance could deviate significantly from the parameterization provided for scientific analysis, so the affected time intervals are flagged as bad.

\subsection{Calibration}\label{sec:calib}

Fully calibrating an instrument the size of the LAT with a beam test is generally not feasible: the full detector is too big for the beam facilities and transport and irradiation are, in any case, too risky. In the case of the LAT, structures and modules were tested with radiation sources at each stage in the production and assembly \citep{tkr-ssd-design,tkr-ssd-rad,cal-rad}, including one full CAL module (known as the ``Engineering Model'', slightly different from the flight models) \citep{cal-ions}. In this section, we focus on the larger assemblies only, containing sensors for all the three LAT subsystems. 

The basic concept of the LAT was validated early in 1997, using a test structure built with simple versions of the planned flight sensors (a few SSDs, a hodoscopic arrangement of CsI  crystals, plastic scintillator tiles for anti-coincidence) \citep{beamtest-slac}. The outcome of the campaign was the validation of the design choices, verifying the expected performance. Most importantly, comparison of the results with the Monte Carlo simulations confirmed that the software tools accurately reproduced the instrument performance. A follow-up beam test of a structure resembling one LAT tower module (the Beam Test Engineering Model, BTEM) was performed in 1999/2000 with similar results \citep{beamtest-slac2}. A balloon flight was performed in 2001 using a detector similar to the BTEM; it confirmed that the LAT design could operate in a space-like background environment \citep{balloon}.

The ``Calibration Unit'' is a scaled down instrument, assembled using the two spare towers (including TKR and CAL), 1 additional CAL module and 5 ACD tiles. The main purpose is to have on ground a platform to perform tests and replicate issues that may happen in space. A final on ground calibration campaign was performed \citep{beamtest-cern} to verify the calibration procedures and to tune the parameters of Monte Carlo simulation: backsplash from the calorimeter, energy leakage corrections, background rejection techniques, etc. 


The calibration procedures, validated via the beam tests on-ground, are routinely carried out in space \citep{onorbitcalib}. Many parameters must be evaluated and finely tuned: gains, thresholds, alignment, time delays, live time, etc. 
Data needed for calibration are constantly collected during normal science operation, in particular thanks to the dedicated trigger engines. \tabref{triggers}\ shows one such example, engine 4, requiring a high energy deposition in the ACD, a track in an associated TKR tower, and some energy deposition in the CAL. This engine is very effective at collecting heavy ions crossing both TKR and CAL. 
In addition, calibration runs in \emph{dedicated mode} are scheduled periodically to collect data that cannot be obtained during nominal operation, as discussed in the section about data acquisition. The LAT spends approximately 2.5~h every 3~months in dedicated mode. 

Monitoring the calibration parameters as a function of time is also an effective way of monitoring the health and stability of the instrument. After more than 12 years of operation the instrument remains in excellent conditions. Let us mention only a few examples, for a detailed discussion see \citet{onorbitcalib,instr10y}.

\begin{figure}[htbp]
  \centering
    \includegraphics[width=0.49\textwidth]{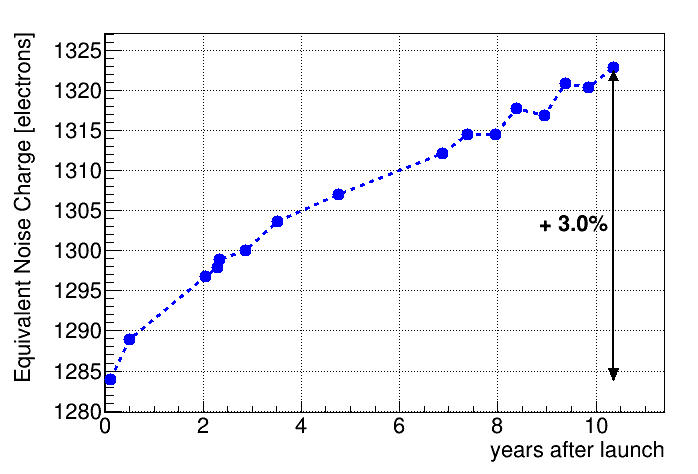}
        \includegraphics[width=0.49\textwidth]{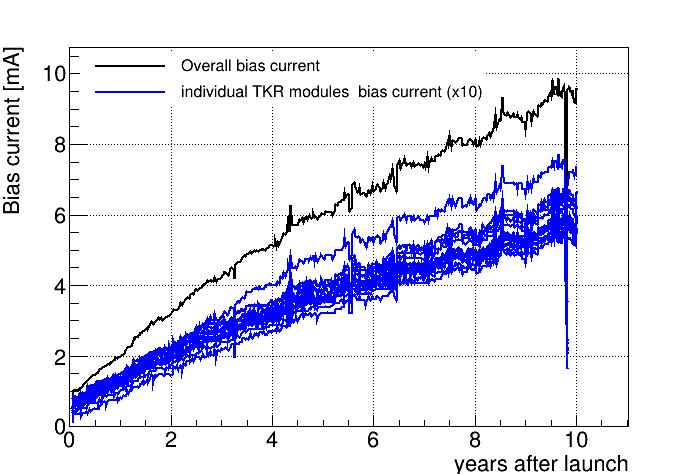}
  \caption{Left: TKR readout noise averaged over all channels. Right: TKR bias current over time, for the 16 tower modules (scaled $10\times$) and total. From \citet{instr10y} \copyright\ AAS.}
  \label{fig:tkrevol}
\end{figure}

The most evident effect related to aging is the increase of power consumption in the TKR. In \figref{tkrevol}, right, the current drawn by the TKR is plotted as a function of time. In the same plot the current for all the 16 tower modules is shown, multiplied by 10 to fit in the same scale. The slow increase is attributed to the expected radiation damage in the SSDs. In \figref{tkrevol}, left, the corresponding increase in the noise level in the TKR readout is shown. While non negligible, the increase has no consequence: the noise has reached $\sim 1325$ equivalent electrons, or $0.21$~fC, to be compared to the average signal for a MIP at $\sim 5$~fC. In particular there is no need to update the TKR noise discriminator threshold from the initial	 values of $1/4$ to $1/3$ of a MIP.

\begin{figure}[htbp]
  \centering
    \includegraphics[width=0.8\textwidth,trim=0 -2cm 0 0]{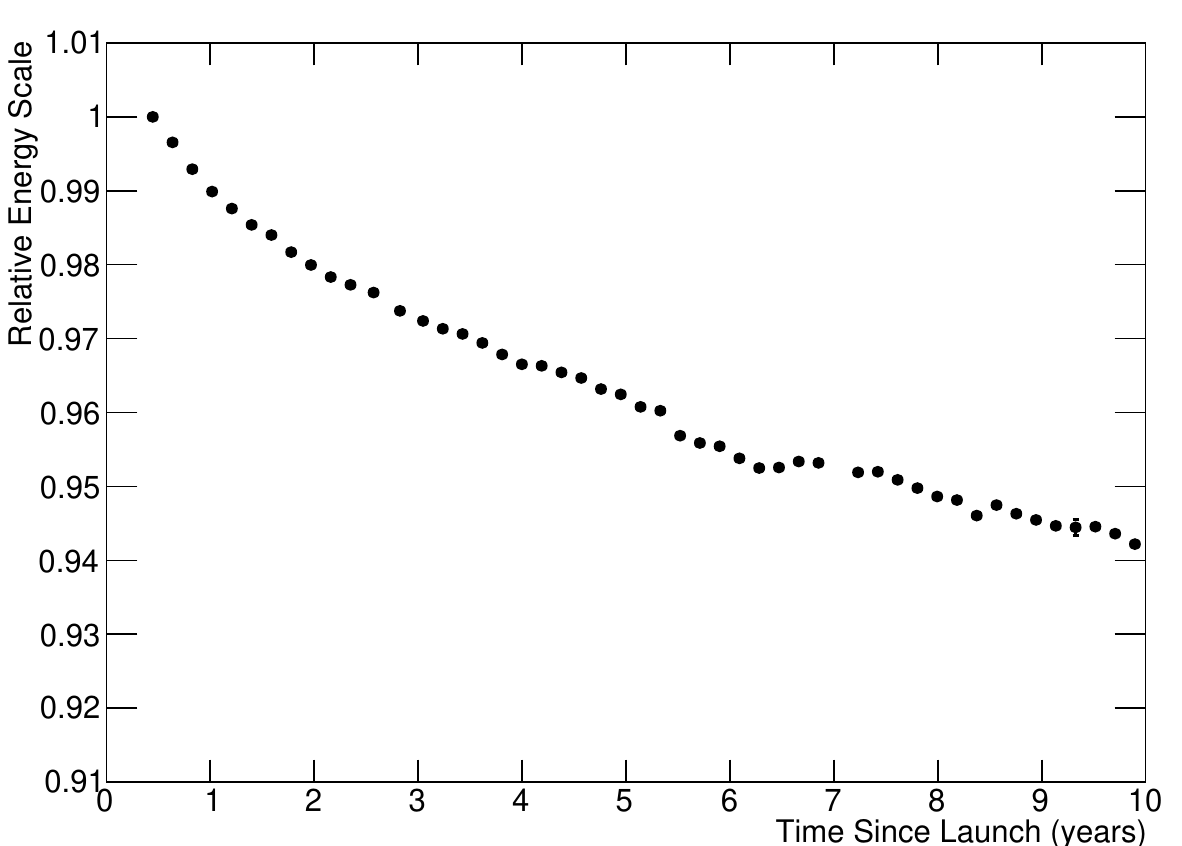}
  \caption{Uncalibrated relative energy scale of the CAL, averaged over all crystals and normalized to the value after launch; from \citet{instr10y} \copyright\ AAS.}
  \label{fig:calevol}
\end{figure}

The second most evident effect is the loss of light yield in the CAL due to radiation exposure of the CsI crystals: a degradation of $\lesssim 1\%$ per year is observed, see \figref{calevol}, and easily managed with a corresponding change in the CAL energy calibration. We mentioned that light absorption in the crystals cause yield loss but improve position resolution. This applies to radiation damage: the light asymmetry along the crystals is increasing very slightly as a function of time, leading to a small improvement in position resolution.

The relative mechanical alignment of the detectors within the LAT affects directly the track reconstruction and the angular resolution, while the alignment between the LAT and the spacecraft affects the conversion between internal LAT coordinates and sky locations. 
Intra-tower and inter-tower TKR alignments use reconstructed events recorded during nominal science operations with no special selection, 
minimizing the possibility of selection bias. Residuals in the track fitting procedure are converted into geometrical displacement of the SSD planes (three traslations and three rotations) and included in the calibration. The sensitivity of the measurement is better than 2~\um\ for translations, and than 0.02 mrad for rotations around the coordinate axes, roughly an order of magnitude below what would affect the angular resolution at high energies. No significant time evolution is observed. 
The \fermi\ reference system is determined by the optical star tracker system in the spacecraft. The relative alignment of the LAT to the spacecraft is determined by
minimizing the residuals of the measured locations of known
\gammaRayHyph\ sources in the sky: an accuracy better than $5^{\prime\prime}$ on the three rotation angles is achieved, again with no significant time evolution.


\subsection{Performance}\label{sec:perf}

An accurate description of the  instrument performance is necessary to reconstruct astrophysical quantities (e.g. \gammaRayHyph\ flux, source extension, \dots) from the observed event counts. The instrument design, reconstruction, and event analysis  determine the overall performance. In particular, candidate \gammaRays\ may be assigned to several different event classes, and a class selection defines an event set and a corresponding performance.

\begin{figure}[htbp]
  \centering
    \includegraphics[width=0.49\textwidth]{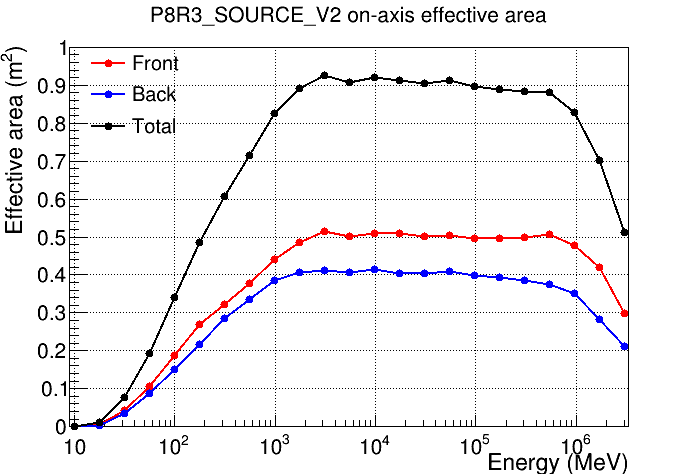}
    \includegraphics[width=0.49\textwidth]{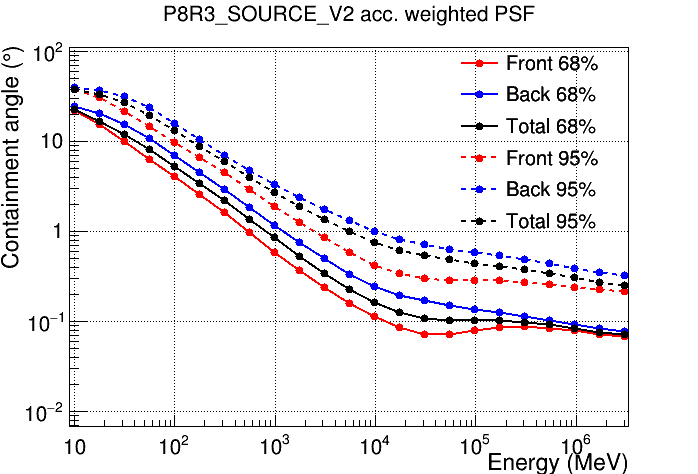}
    \\
    \includegraphics[width=0.49\textwidth]{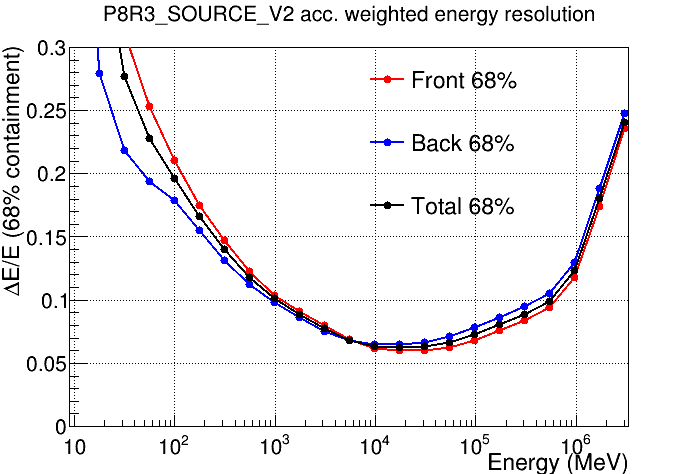}
    \includegraphics[width=0.49\textwidth]{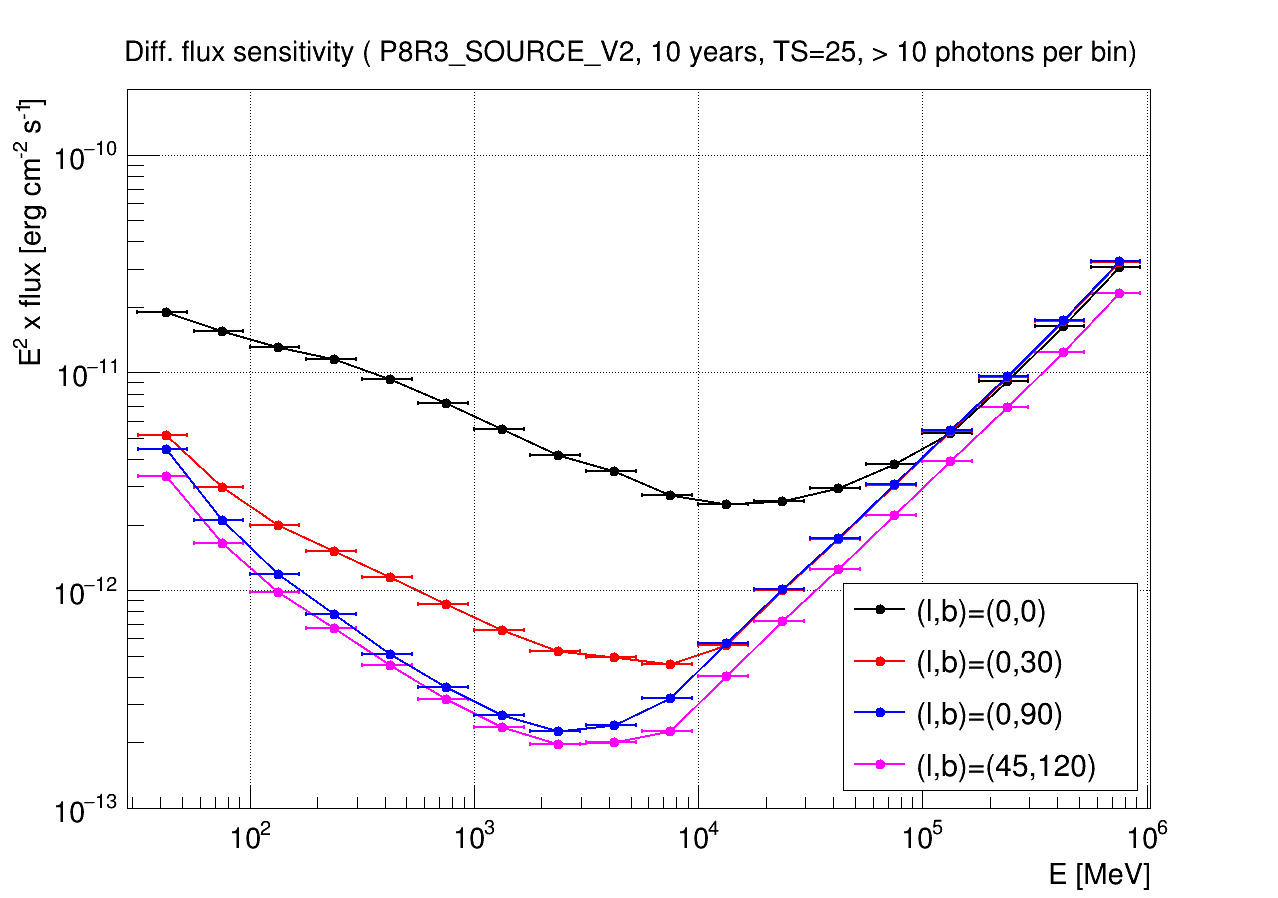}
    
  \caption{Performance plots, from \citet{perf-site}. See the text for a discussion.}
  \label{fig:perfplots}
\end{figure}

The \emph{Instrument Response Function} (IRF) is the map converting the incoming photon flux into detected events; the IRF is canonically factorized into \emph{effective area} (i.e. geometric area times efficiency), \emph{point spread function} (angular resolution) and \emph{energy dispersion} (energy resolution). For each event class, the corresponding quantities, as a function of the \emph{true} energy and direction in the instrument frame of reference, are evaluated with Monte Carlo simulations. Since launch, such simulations have been tuned to best replicate the behavior and quirks observed in real data \citep{p7paper}. 

An example of LAT performance plots is shown in \figref{perfplots}. The name of the event class (``\techname{P8R3\_SOURCE\_V2}'') indicates that this refers to the current \Peight\ event analysis, release 3. The ``\techname{SOURCE}'' event class is tuned for the analysis of non-transient point sources; other event classes are available, optimized for the study of transients, diffuse components, etc. The final version number refers to the IRF parametrization, the progressive version number includes test versions and release candidates so not all versions are released publicly.

The top left, top right, and bottom left plots in \figref{perfplots} show example values of effective area, point spread function, and energy resolution, respectively. The events converting in the \emph{front} and \emph{back} part of the TKR are shown separately, in addition to the total, or average, value. All figures of merit are described in terms of \emph{true} (i.e. Monte Carlo) energy and inclination bins. The first plot shows the effective area as a function of energy for on-axis incident \gammaRays; an additional (few \%) dependence on the azimuthal angle around the LAT axis, due to the square shape and the alignment of the gaps parallel to the sides of the instrument, is averaged out to produce the plot. The remaining plots account for the dependence on the off-axis angle, by integrating the effective area over the solid angle.
The details of the IRF implementation and the procedure to generate the plots above are described in great detail in \citet{p7paper}.

The bottom right plot in \figref{perfplots} shows the sensitivity for point source detection at different locations in the sky, derived semi-analytically from the instrument performance and from a background model (10 years of observation, $5\sigma$ sensitivity) \citep{p7paper,perf-site}.

\subsection{Conclusion}\label{sec:end}
With no consumables limiting its lifetime, after more than 13 years of operation, the LAT remains in excellent operating condition.
Considering the continuing good performance, the LAT can play a major role in the new era of multi-messenger astrophysics alongside the existing and future instruments \citep{neutrinos,gravitational,cta,rubin}. 

Since the time of the LAT design phase, no game-changing new technology in the field has appeared, at least nothing comparable with the shift from gas spark chambers to Silicon microstrip trackers. An important, but not revolutionary, improvement in detector technology is the appearance of Silicon photomultipliers (SiPM) as a replacement for photomultiplier tubes \citep{sipm}, and the good performance of scintillating fiber trackers is also worth mentioning \citep{fibers}. All things considered, the design of the LAT remains very close to the state of the art for imaging \gammaRayHyph\  observatories. A different optimization can be sought, e.g. a thicker calorimeter for a better performance at high energy, at the cost of a smaller FOV, see e.g. the \gammaRayHyph\ performance of CALET \citep{calet}. On the other hand, the heritage of the LAT is evident in the design of the proposed future space observatories in the MeV regime \citep{amego,astrogam}.

Until a new breakthrough in HEP detectors occurs, the LAT will remain the best all-purpose, wide FOV \gammaRayHyph\ instrument covering the energy range from the onset of pair production at a few tens of MeV to a few hundred GeV energies with excellent performance, delivering invaluable scientific data.

\bibliographystyle{spbasic}
\section{Cross-references}\label{sec:crossref}

\begin{enumerate}
\item[] Thompson, D and Wilson-Hodge, CA  (2021) Fermi Gamma-ray Space Telescope, in this volume. 
\end{enumerate}

\bibliography{fermilat}

\end{document}